\documentclass{mn2e}
\usepackage{epsfig}
\usepackage{pslatex}
\usepackage{amssymb}

\def\msun{{\rm M_{\odot}}}

\def\today{\number\year \ \ifcase\month\or
  January\or February\or March\or April\or May\or June\or
  July\or August\or September\or October\or November\or December
 \fi \ \number\day }
\date{Accepted ??. Received ??; in original form \today}

\volume{000}

\pagerange{\pageref{firstpage}--\pageref{lastpage}} \pubyear{2006}

\begin{document}

\label{firstpage}

\title[On Rejuvenation in Massive Binary Systems]{On Rejuvenation in Massive Binary Systems}
\author[L. M. Dray \& C. A. Tout]
       {Lynnette M. Dray$^{1}$\thanks{E-mail: Lynnette.Dray@astro.le.ac.uk} and Christopher A. Tout$^{2}$
\\ 1. Department of Physics and Astronomy, University of Leicester, University Road Leicester, LE1 7RH, UK.
\\ 2. Institute of Astronomy, Madingley Road, Cambridge, CB3 0HA, UK.}

\date{Accepted:
      Received:}
\maketitle

\begin{abstract}

We introduce a set of stellar models for massive stars whose evolution has been affected by mass transfer in 
a binary system, at a range of metallicities.
As noted by other authors, the effect of such mass transfer is frequently more than just 
rejuvenation. We find that, whilst stars with convective cores which have accreted only H-rich matter rejuvenate 
as expected, those stars which have accreted He-rich matter (for example at the end stages of conservative mass
transfer) evolve in a way that is qualitatively similar to rejuvenated stars of much higher metallicity.
Thus the effects of non-conservative evolution depend strongly on whether He-rich matter is amongst the 
portion accreted or ejected. This may lead to a significant divergence in binary evolution paths with only a small 
difference in initial assumptions. We compare our models to observed systems and find approximate formulae for 
the effect of mass accretion on the effective age and metallicity of the resulting star.

\end{abstract}


\section{Introduction}

The effects of binary evolution on massive stars are many and complex. In the simplest case, the two stars do 
not interact at all; wide systems which evolve essentially as two single stars make up about half of all binaries.
In closer systems, the stars affect each other via the interaction of their winds, via tidal interaction and, 
most importantly for their later evolution, via Roche lobe overflow (RLOF). Whilst there are many massive binaries which 
can be surmised to have gone through this phase (e.g. Podsiadlowski, Joss \& Hsu 1992; Wellstein \& Langer 1999), the 
amounts of mass and angular momentum transfer are frequently rather poorly-constrained and appear to vary between 
systems (Langer et al. 2003; Petrovic, Langer \& van der Hucht 2005). In turn, the amount of 
mass and angular momentum transfer determines whether the stars enter a common envelope phase (Webbink 1984; Beer et 
al. 2006), and this has a strong effect on the resulting masses and period of the system at the end of interaction, 
if the two stars do not merge. The current parameters of observable massive binaries are vital in pinning down the 
many uncertainties about the mass transfer process. Models for massive binary evolution over wide ranges of
parameter space and physical complexity have been computed by several 
groups (e.g. Nelson \& Eggleton 2001; Petrovic et al. 2005; Podsiadlowski et al. 1992; de Donder \& Vanbeveren 2004) 
although most concentrate only on solar metallicity (for an exception see de Loore \& Vanbeveren 1994). 
The general consensus seems to be 
that several different mass transfer mechanisms must be responsible: for example, some systems require strongly non-conservative 
mass transfer (Clark et al. 2002), whereas some require at least quasi-conservative mass transfer to reach their current parameters from a 
plausible starting point (Wellstein \& Langer 1999; van Rensbergen, de Loore \& Vanbeveren 2005). 

The outcomes of binary evolution are also important to models of massive star populations. Observations suggest that 
the cluster O-star binary fraction is high, perhaps greater than 75 per cent and maybe approaching 100 per cent when hard-to-observe areas of the parameter 
space are considered (Mason et al. 1998). B star binary fractions are also large (Raboud 1996). Given that it is much 
easier to unbind a binary than to create one from two initially single stars, this suggests that nearly all massive 
stars have at least one companion at formation. A high initial binary fraction is also suggested by 
the some models of star formation (e.g. Delgado-Donate et al. 2004) and may be required to 
reproduce the observed number of stars with runaway velocities (de Donder, Vanbeveren \& van Bever 1997, Dray et al. 2005). If this is the case, 
then many single massive stars in late stages of evolution may have had a binary past (Vanbeveren et al. 1998).
The output of massive binary simulations is therefore also of much use as an input to population synthesis models 
(e.g. Hurley, Tout \& Pols 2002; O'Shaughnessy, Kalogera \& Belczynski 2006; Podsiadlowski et al. 2004). 
Here, one must make a decision as to which approximations to use in describing the evolution of binaries, because to use full evolutionary tracks 
can take prohibitively long.

\begin{figure*}
\vbox to75mm{\vfil
\psfig{figure=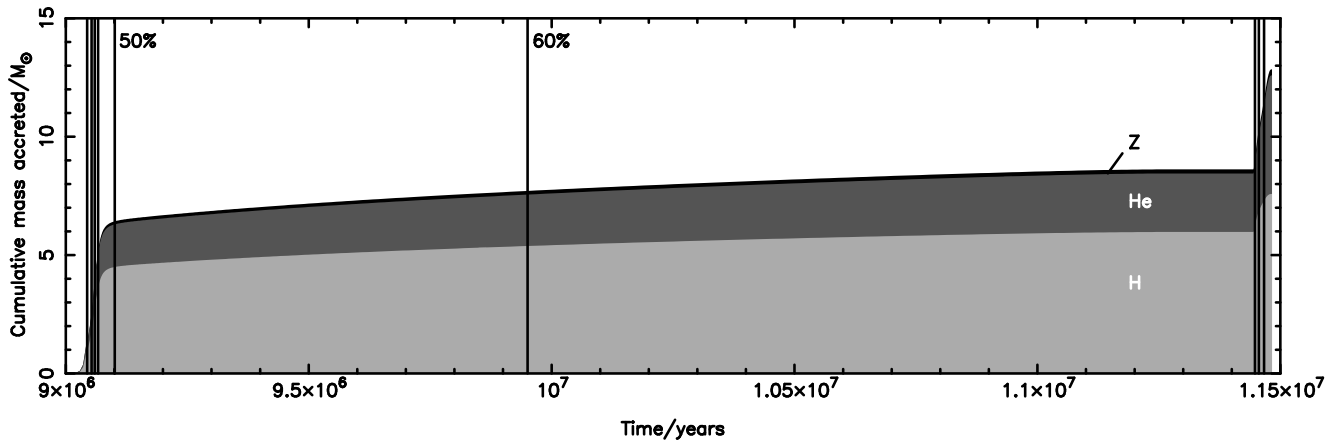,angle=270,width=175mm}
\caption{Amounts of different elements accreted with time for a typical sequence of conservative mass transfer episodes. 
It is notable that 
whilst the metallicity of the accreted matter remains essentially unchanged, the amount of helium accreted goes up rapidly 
during the final stages of mass transfer. Also shown (black lines) are the time intervals for each 10\% of the mass to be accreted.
This system has initial masses of $16.6$ and $15.7 \msun$ and an initial period of three days, i.e. mass transfer is initially 
type A.} 
\vfil}
\label{fig1}
\end{figure*}

For both investigation of individual massive stars and whole-population properties, then, the behaviour of the secondary of a binary system 
after it has accreted mass is at least of interest and may be of vital importance. 
Previous investigations of this behaviour (Braun \& Langer 1995; Vanbeveren \& de Loore 1994) have suggested that the usual 
assumption about the evolution 
of such stars, that they are rejuvenated by some amount (Hellings 1983) and thereafter evolve similarly to 
a younger star of their new mass, is true in some cases but not all the time. In particular, mass accretion on to a massive main-sequence 
star, even if it is of the same composition as the surface of the star, causes the convective core to expand, leading to a molecular 
weight discontinuity at the core boundary. Braun \& Langer (1995) did not consider the accretion of He-enhanced matter but it is likely that
this too may have a strong effect. Such enriched matter is likely to be transferred in the last stages of conservative mass transfer, or in later 
mass-transfer stages after an initial one (e.g. case AB mass transfer), when the core of the donor star has been nearly exposed.
When matter of a higher molecular weight is accreted on top of matter of a lower molecular weight, the 
thermohaline instability occurs (Kippenhahn, Ruschenplatt \& Thomas 1980). Analagously to the more commonly-known thermohaline mixing process 
in oceanography, this instability allows mixing downwards of the accreted matter to occur locally even if overall the  
criteria for convective stability are met. This effect has also been investigated for lower-mass stars by Chen \& Han (2004) in 
the context of blue straggler populations. 

In particular, previous investigations (Dray \& Tout 2006) have suggested to us that the threshold mass above which a
star goes through a Wolf-Rayet (WR) phase at the end of its lifetime is substantially reduced for stars which have accreted He-rich matter.
Wolf-Rayet stars are the H- and He-depleted cores of the initially most massive stars, stripped of their 
envelopes by high levels of mass loss, and as such are the likely progenitors of long gamma-ray bursts (GRBs).
Originally (Paczy{\'n}ski 1973) it was thought that this mass 
loss was a result of Roche lobe overflow in a binary system but more recent models have suggested that
the vast majority of WR stars can arise without any form of interaction with other stars, via rotationally-enhanced 
wind mass loss only (Maeder \& Meynet 2000). However, the large binary fraction amongst O stars, which (at least in the single 
star scenario) are the sole progenitors of WR stars suggests that duplicity is important. If the threshold mass for WR formation is 
substantially lowered amongst secondaries it could have effects on both the CNO enrichment from the system and the ratio of type-II
to type-Ibc supernovae. This may be particularly important at low metallicity where it is hard to form a WR-type star via stellar winds 
only. In this paper we investigate this behaviour 
in further detail, with a particular emphasis on finding approximations to the behaviour of these rejuvenated stars which 
can be used in population synthesis models.

\section{Models}
Modelling massive binary stars is a pursuit hampered by the massive parameter space involved. This parameter space 
is at least three-dimensional, since one must consider the initial primary mass, the mass ratio and the initial period 
to get reasonable coverage of a population. Added to these is the uncertainty inherent in the modelling of single massive stars -- 
in particular, the mass-loss rates and any coverage of rotation. We are interested in the effects of metallicity 
and the amount of mass transferred, a further two variables. In previous work (e.g. Dray \& Tout 2005) we have dealt 
with this complexity by running large grids of models using the {\sc stars} code (Eggleton 1971; Pols et al. 1995, 1998 and references therein) 
in a quasi-simultaneous mode (see e.g. Pols 1994), and then checking the accuracy of edge cases against full models 
using the more recent fully simultaneous version of the code, {\sc twin} (Nelson \& Eggleton 2001; Eggleton \& Kiseleva-Eggleton 2001). 
However, in this paper we are more interested in the detailed behaviour of accreting stars rather than the specific fate of binaries 
containing them (e.g. the period evolution) or integrated whole-population properties. We therefore run smaller grids of models 
using only the {\sc twin} code, allowing for a reasonable range of metallicities and initial masses but not covering the entire 
range of different combinations. As is apparent below, the evolution of the accreting stars that we simulate appears to be quite 
straightforwardly generalisable to accreting stars in general. In particular we focus our attention on two series of models, 
18 x 9 grids of mass ratio and period for an initial primary mass of $16.6 \msun$ over a range of metallicities from $0.03$ to 
$0.0001$ and 14 x 9 grids of mass ratio and period for solar metallicity and initial primary mass between 10 and $20\,\msun$. This 
covers in detail the mass range over which it is possible that conservatively-accreting secondaries reach the threshold mass for 
WR evolution, where we have previously found interesting behaviour with less-detailed models. Over the mass range around the WR 
threshold the systems which avoid common envelope evolution are in the case A (mass transfer begins whilst the donor is undergoing 
core hydrogen burning) and early case B (mass transfer begins shortly after this) regimes, so we have 
concentrated on this area of the parameter space. We have also run a number of other grids 
at different metallicities for the purposes of testing the generalisations about the behaviour of accretion stars formed from 
these models.

\begin{figure}
\vbox to145mm{\vfil
\psfig{figure=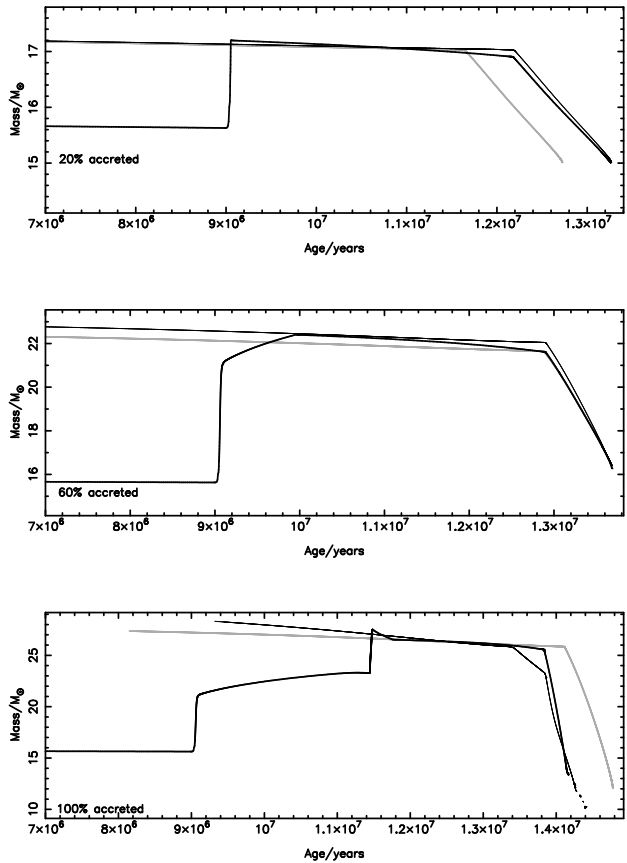,angle=0,width=82mm}
\caption{Evolutionary tracks for some of the non-conservative accreting secondaries whose interaction with a binary companion was
cut off at the points shown in Fig. 1. Shown are the secondary's mass with age (thick black line), the closest matching model and 
time offset according to a simple rejuvenation recipe (Tout et al. 1997, grey line) and the closest matching model and time offset 
from fitting to a large library of stellar tracks at multiple metallicities (thin black line). Dotted lines indicate that the star 
is in a WR phase. It is notable that the simple recipe fits remarkably well for accretion of matter which is not helium enriched.}
\vfil}
\label{fig2}
\end{figure}
 
Unless otherwise specified, the physical ingredients used in the code are those used in our previous work (e.g. Dray \& Tout 2003)
or which are standard in the {\sc stars} and {\sc twin} code (e.g. Pols et al. 1998), including the reaction rates, opacity tables 
and convective overshooting prescriptions; 
mass-loss rates for WR stars are taken from Nugis \& Lamers (2000), with mass loss for earlier stages calculated using the 
prescription of Vink et al. (2001) where applicable and otherwise the empirical mass-loss rates of de Jager (1988). We set the 
composition of the accreted matter during each accretion timestep to be equal to the surface composition of the donor star. In many 
cases, particularly if the accreted matter is helium-rich, this leads to significant mixing.  We treat thermohaline mixing from 
accretion of matter of a higher molecular weight than the surface matter in a time-dependent 
way as done for the {\sc stars} code by Chen \& Han (2004), using the mixing timescale of Kippenhahn et al. (1980). In some cases we find 
this leads to numerical instabilities which can be eliminated by capping the mixing rate to exclude non-physical values and suppressing 
mixing when the composition difference is infinitesimal. In cases where the core is convective and accretion leads to a significant amount 
of helium-enriched matter being deposited on the surface of the star, this can essentially lead to mixing of the entire star, with the 
central hydrogen abundance increased and a consequent lengthening (rejuvenation) of the main-sequence lifetime in comparison to 
what is expected for a star of the new mass (Braun \& Langer 1995).
As in Chen \& Han (2004), we also assume that the accretion stream impacts the 
surface of the accretor with zero falling velocity and, as necessitated by the use of a one-dimensional code, is deposited in a 
homogeneous layer over the surface of the star. 

We do not consider the effects of rotation or magnetic fields in our models. Both of these may have a strong 
effect on the evolution (Petrovic et al. 2005; Heger, Woosley \& Spruit 2005), particularly in tandem with binary interaction via the 
accretion of angular momentum; however, there are still significant uncertainties, particularly in regard to the effect of 
magnetic fields, and some models of magnetised, rotating stars evolve more 
similarly to non-magnetised, non-rotating stars than to stars with rotation alone (Maeder \& Meynet 2004). 
Rotation may also limit the amount of matter which can be accreted (Packet 1981; see also section 3).

\begin{table*}
\begin{minipage}{\textwidth}
\centering
  \caption{`Rejuvenated' closest equivalent single star tracks to non-conservative secondaries, for the 
system shown in Fig. 1. Note that the matching models are taken from a library of stellar tracks without 
interpolation, so there is an uncertainty of a few tenths of a solar mass associated with them.}
  \begin{tabular}{@{}rrrrr@{}}
 {\bf Amount of}& {\bf Secondary mass at}& {\bf Matching single} & {\bf Matching single star} & {\bf Matching single}\\ 
 {\bf matter accepted}& {\bf end of transfer}& {\bf star model} & {\bf time offset/years} & {\bf star metallicity} 
\\ 
\hline
10 \% & $16.9 \msun$ & $16.5 \msun$ & $8.4 \times 10^{6}$ & $0.02$\\
20 \% & $17.2 \msun$ & $17.5 \msun$ & $7.3 \times 10^{6}$ & $0.02$\\
30 \% & $18.5 \msun$ & $18.6 \msun$ & $5.4 \times 10^{6}$ & $0.02$\\
40 \% & $19.8 \msun$ & $19.9 \msun$ & $5.3 \times 10^{6}$ & $0.02$\\
50 \% & $21.1 \msun$ & $21.5 \msun$ & $4.5 \times 10^{6}$ & $0.02$\\
60 \% & $22.4 \msun$ & $22.9 \msun$ & $4.7 \times 10^{6}$ & $0.02$\\
70 \% & $23.7 \msun$ & $24.6 \msun$ & $4.8 \times 10^{6}$ & $0.03$\\
80 \% & $24.9 \msun$ & $26.1 \msun$ & $4.2 \times 10^{6}$ & $0.03$\\
90 \% & $26.2 \msun$ & $27.2 \msun$ & $2.9 \times 10^{6}$ & $0.05$\\
100 \% & $27.5 \msun$ & $28.3 \msun$ & $2.6 \times 10^{6}$ & $0.05$\\
\hline
\end{tabular}
\end{minipage}
\end{table*}

Angular momentum loss via winds is treated as in Hurley et al. (2000), as is the potential accretion of wind matter from one star by the 
other via the Bondi-Hoyle process. Wind accretion is not applied if both winds are strong, as expressed in terms of wind momentum
(Walder \& Folini 2000); 
here we would expect a colliding wind system instead. Whilst we consider a form of non-conservative accretion in the 
following section, we do so in order to look at the behaviour of the accreting star if there is no further interaction 
in the system. For those purposes we treat it as a single star and there is no need to specify a mode of angular momentum loss during 
Roche Lobe overflow. If the two stars evolve into contact we halt the evolution of the system and do not follow it any further. In many 
schemes for contact and common-envelope evolution the secondary accretes no further matter after the onset of this phase (e.g. Webbink 1984)
and in that case its evolution should be similar to that of a star which has accreted only a portion of the available matter, as with 
rapidly-rotating stars. However, once again, significant uncertainties are involved in the common-envelope phase which make it difficult to 
follow.

\section{Evolution with amount transferred}
Before looking at full conservative mass transfer sequences, it is interesting to follow how the amount of accretion affects
the subsequent evolution. It is likely that most accretion events are non-conservative, not least because of the effects of accreting 
angular momentum (e.g Packet 1981). Therefore if there is a significant difference in the subsequent behaviour of accreting stars 
with the amount of matter they are able to accrete, this may lead to systems following quite different evolutionary paths depending on 
their accretion history. 

Another consideration for stars which accrete only a portion of the transferred matter is {\it which} portion they accrete. At least 
in the case of the initial occurrence of RLOF in the binary, the material which is first transferred will be at the same composition as 
the accreting star. Thermohaline mixing is therefore a relatively minor consideration. However, later on in the same accretion episode 
the donor may have been stripped down to the helium-enhanced regions near its core, leaving the transferred matter significantly 
helium-enriched (Fig. 1). If this matter is accreted, deep mixing (potentially of the entire star if the thermohaline mixing region meets the 
convective core of the accretor) is likely. Therefore if one assumes non-conservative accretion in which (say) ten percent of the transferred 
matter is accreted, the resulting evolution may differ significantly depending upon whether this accepted ten percent is the first ten percent of 
transferred matter (thereafter no further matter being accepted) or if initially and throughout only ten percent of the matter 
transferred at any given time is accreted. The former is the more likely case (e.g. Packet 1981; Dewi 2006).

To evaluate the effect of the amount of accreted mass in the former case, we have run some example systems in which binary evolution 
is halted at a number of points throughout a conservative accretion event (see marked lines in Fig. 1 above) and the subsequent evolution 
of the accretor is carried out with the code in a single star mode with no further accretion. This mimics non-conservative accretion of 
varying proportions of 
the transferred matter in systems with no further interaction. It is commonly assumed that the secondary star in 
a common envelope phase accretes no further matter from the primary, so these calculations may also be relevant there as well (although 
systems which survive common envelope are likely to remain close enough that further interaction is likely). In Fig. 2 we show some 
evolutionary tracks from secondaries which have accreted differing amounts of matter from the system shown in Fig. 1. We also perform 
least-squares fitting against a detailed library of single-star evolutionary tracks with the same physical ingredients 
(updated from Dray \& Tout 2003)
to determine the closest rejuvenated match for the evolution and  
evolutionary type of these stars, which are detailed in table 1\footnote{Note that fitting is not carried out against the star's luminosity; 
in general we find that accretion stars are overluminous in comparison with models which otherwise fit the mass, type and effective 
temperature evolution with time well. As noted by Braun \& Langer (1995), it is often not possible to fully match the evolution 
of an accretion star to a single star model because the internal chemical structures differ.}. 

As can be seen, for most forms of non-conservative mass transfer 
during which only early-transferred matter is accreted, straightforward rejuvenation is the best fit. Various approximate formulae 
for the effects of rejuvenation exist; perhaps the most basic is that after an accretion star reaches its maximum mass, it becomes a 
ZAMS star of the new mass (e.g. Portegies Zwart 2000). A slightly more complex version is one in which the accretion star is assumed 
to behave like a star of the new mass, but offset from the ZAMS by some amount of time which is governed by (an approximation to) 
the core hydrogen abundance increase (hence the increase in core H-burning lifetime). We use the simple formula 
from Tout et al. (1997), 
\begin{equation}
t' = \frac{M}{M'}\,\frac{\tau_{\rm MS}'}{\tau_{\rm MS}}\, t ,
\end{equation}
where $t$ is the age of the secondary at the time of maximum mass\footnote{This time may be substantially different from the time 
of the initial onset of RLOF -- in particular, for case A systems, we can find a later time of maximum mass with shorter period, 
whereas RLOF always starts earlier with a shorter period. However, the bulk of the helium accretion is likely to occur towards the end of 
the accretion period.}, $t'$ is the effective age after rejuvenation, $M$ and $M'$ are the initial and post-RLOF
masses of the secondary (assumed for these stars to be proportional to the remaining fraction of unburnt hydrogen in the convective 
core, see Tout et al. 1997 for details) and $\tau_{\rm MS}$ and $\tau_{\rm MS}'$ 
are the main-sequence lifetimes of single stars at the old and new masses of the secondary. For the main-sequence lifetimes we also use
their fitting formulae, which depend only on mass. This gives us an effective time offset, which can compared with the 
best-fit model
and depends only on the initial mass, age and amount accreted by the secondary. As is indicated in Fig. 2, this approximate formula
produces good results here, provided that the accreted matter is similar in composition to the surface of the accreting star.  
However, for nearly or wholly conservative 
mass transfer we find a better fit to tracks for stars close to the new rejuvenated mass, but of a higher metallicity. This holds 
true even though the effective metallicity of the new star has not increased. In some respects this behaviour is not particularly 
surprising, as it is expected both from theory and observation that stars which have accreted a lot of helium will have raised 
surface helium abundances (Blaauw 1993, Vanbeveren \& de Loore 1994)
and the surface abundances have an effect on the assumed mass-loss rates and hence the evolution. However, this similarity 
to higher-metallicity evolution continues throughout the star's lifetime and persists even though the star remains overluminous 
for its type, which should also affect the mass-loss rates. In panel 3 of Fig. 2, for example, the accretor 
undergoes a WR phase at the end of its lifetime. Although this phase is short, it is important both in terms of the 
potential chemical enrichment from the star (as there are high mass-loss rates combined with unusual surface abundances) and 
the type of final explosion one would expect to see (this model produces a type Ib supernova, whereas the 
closest same-metallicity match, shown in grey, would produce a type II supernova). Whilst the effective metallicity of this star remains 
the same, if one defines a 'helium metallicity' -- that is, the metallicity one 
might expect to see in a main-sequence star with a particular helium abundance, assuming a simple scaling from solar metallicity -- 
this seems to be a reasonably accurate predictor of the metallicity which the evolutionary tracks mimic.
For the input models we have used, the 
initial helium Y is scaled with metallicity Z as ${\rm Y = 0.24 + 2 Z}$, giving a helium metallicity
\begin{figure}
\vbox to152mm{\vfil
\psfig{figure=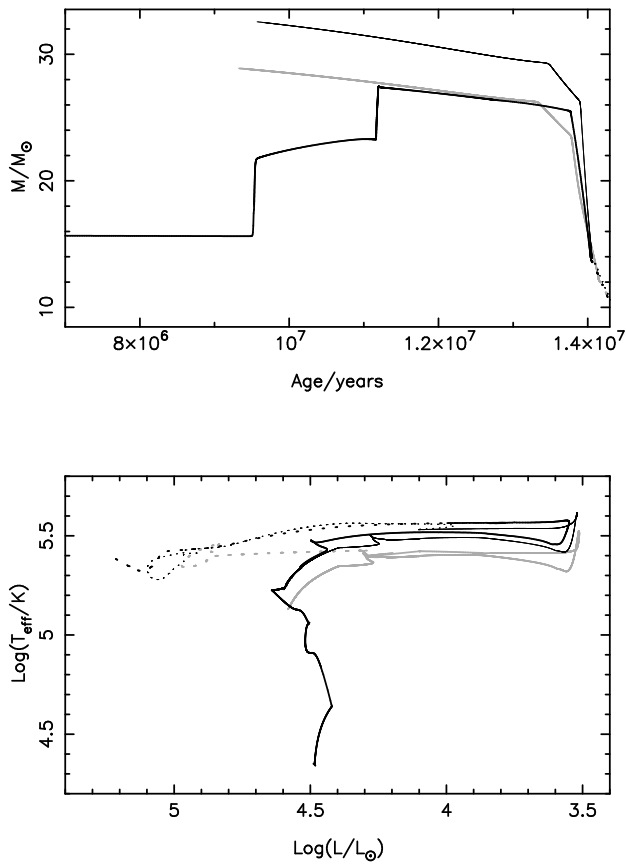,angle=0,width=82mm}
\caption{Comparison of single stars tracks which provide a good fit in terms of mass 
evolution and HR diagram position evolution (all are fitted to evolutionary stage). Thick 
black lines show the mass evolution (upper panel) HR diagram track (lower panel) for 
the secondary in a solar-metallicity conservatively-accreting binary with initial masses $16.6$ and $15.7 \msun$ 
and initial period $3.5$ days. Thin black lines show a model which is a reasonable match to 
the HR diagram position with time (initial metallicity $0.04$, initial mass $32.5 \msun$), 
and grey lines a model which matches the mass evolution (initial metallicity $0.05$, initial 
mass $28.9 \msun$). As noted previously, there is no one single star model which provides a good fit to 
all quantities.}
\vfil}
\label{fig3}
\end{figure}
\begin{equation}
{\rm Z_{He} = (Y + M_{He,acc}/M_{2})/(2 + 2\,M_{acc}/M_{2}) - 0.12} \,, 
\end{equation}
where Y is the initial helium abundance by mass fraction, ${\rm M_{2}}$ is the initial mass of the accretor, 
${\rm M_{acc}}$ is the total amount of accreted matter
and ${\rm M_{He, acc}}$ is the mass of helium accreted. Of course in real stars the helium and metal abundances, although correlated, are not 
dependent quantities, so a further degree of freedom is possible here; and in addition, since mass transfer generally happens after the zero-age 
main sequence, the accreting star will have synthesised some helium of its own. However, the helium metallicity calculated in this way seems to 
be a reasonable predictor of the behaviour of an accretion star. It should also be noted that the above formula is of course specific to the 
initial element mix used in the input models to the {\sc stars} code and therefore the precise calibration to the results of other codes may vary.

\begin{figure*}
\vbox to140mm{\vfil
\psfig{figure=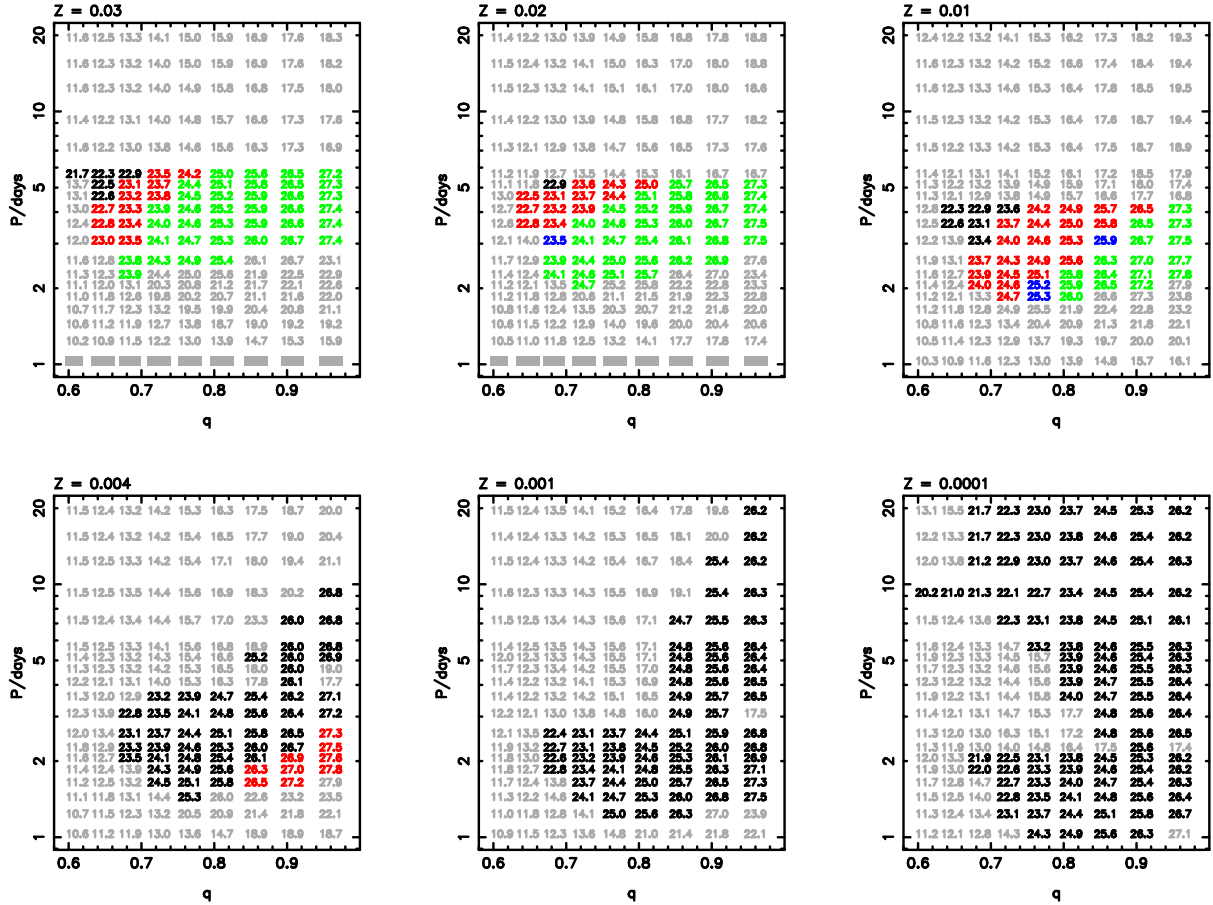,angle=270,width=160mm}
\caption{Initial period -- initial mass ratio diagrams showing the fate of systems with initially $16.6 \msun$ 
primaries at varying metallicity. The numbers shown at each point are the maximum masses attained 
by the secondary. Systems shown in grey go through a contact or common envelope phase. Those in black 
avoid contact and the secondary evolves through to the end of its life without going through a Wolf-Rayet 
phase; systems shown in red, blue and green end their lives in WNL, WNE and WC phases respectively. 
The effect of decreasing stellar radii with decreasing metallicity means that the behaviour of 
low-metallicity systems is similar to those at higher metallicity but lower mass, but with a 
lower-period boundary between case-A and case-B mass transfer systems. Grey boxes indicate systems
which would undergo RLOF at or soon after formation.}
\vfil}
\label{fig4}
\end{figure*}

Note that this reasoning applies only to stars which have a convective core at the time of mass transfer. In the case of the parameter 
space looked at in this paper (massive binaries with $q > 0.5$ and case-A or early case-B mass transfer) this is true for the vast 
majority of accretors. The time offset calculated with the formulae of Tout et al. (1997) is not accurate for the stars which 
accrete significant amounts of helium, though. Replacing the approximate main-sequence time $\tau_{\rm MS}$ formula with the metallicity-dependent one 
from Hurley, Pols \& Tout (2000) and including the higher effective metallicity of the rejuvenated star produces better agreement, 
although still only within around 5 percent of the best-fit offset. As this expression is rather complex (although it still depends 
only on the mass and metallicity), we do not reproduce it here. As noted 
by Izzard et al. (2006), finding approximations to Wolf-Rayet evolution is rather difficult and therefore this level of accuracy is 
in fact quite good. 

\section{Behaviour with masses and metallicity}

The behaviour of accretion stars in conservative mass transfer binaries is, unsurprisingly, similar to the 90\% and 100\% mass transfer
cases discussed above. As an extremely rough guide, the subsequent evolution of a star which has accreted all the matter supplied to it 
(in a system which avoids common envelope, at least) is similar to that of a single star, initially a few tenths of a solar mass above the 
maximum mass the accretion star reaches and a factor of a few greater in metallicity. For example, many secondaries from binaries at a metallicity 
of 0.004 behave similarly to stars at solar metallicity in their later evolution. However the comparison is not completely exact; the new star 
remains overluminous throughout its lifetime (for example Fig. 3) and hence the single star which is a best fit to the HR diagram position 
of a secondary is not generally the same single star which is a best fit to the mass evolution. Since the HR diagram position is not 
particularly meaningful for WR models which are not coupled with detailed atmospheric models, because their vigorous mass loss results in 
ill-defined outer radii, we concentrate on models which are a good fit to the evolutionary type and mass distribution. As in the example 
above, this behaviour leads to many more secondaries ending their lives in a WR phase than otherwise expected.

\subsection{Mass}
The effect of initial masses on the outcome of binary evolution, and in particular the range of masses which avoid contact at 
solar metallicity, has already been studied by e.g. Pols (1994) and Wellstein, Langer \& Braun (2001) and we refer the interested 
reader to those papers for more detailed discussion. We find good agreement with 
their findings, in particular with regard to the parameter space of contact avoidance (e.g. Fig. 5 in Pols (1994) and Fig. 13 in 
Wellstein et al. (2001) vs. the second panel of Fig. 4 and the sixth panel of Fig. 5 in this paper). As found by those authors, 
the parameter space in which there is stable case-B mass transfer vanishes as initial primary mass increases, leaving only case-A 
systems able to avoid contact above some initial primary mass (in our case around $15\,\msun$\footnote{Of course, if mass transfer is not 
conservative then the area of contact avoidance increases significantly (Dray \& Tout 2005).}). This puts relatively stringent constraints
on the initial parameters of systems which appear to have evolved conservatively (e.g. Wellstein \& Langer 1999). The area of contact-free 
evolution, in general, is bounded on the upper edge of case B by systems in which the secondary expands rapidly enough during accretion 
to come into contact and on the lower edge of case A by systems which survive an initial phase of stable mass transfer but in which 
the evolution of the secondary overtakes that of the primary and leads to unstable reverse mass transfer. As we adopt a grid 
to cover the parameter space rather than a distribution of initial parameters designed to explore the edge cases, we also find
a small area in period just above the case A/case B boundary in which only a small range of mass ratios are contact-free (e.g. 
$P_{\rm i} = 5$ days in the third panel of Fig. 5) and some $q = 0.95$ systems evolve into contact. For these systems, the 
secondary's evolutionary stage is close enough to that of the primary that it is already nearly filling its Roche Lobe when 
mass transfer is initiated, i.e. even relatively moderate mass transfer produces a contact situation. 

\begin{figure*}
\vbox to130mm{\vfil
\psfig{figure=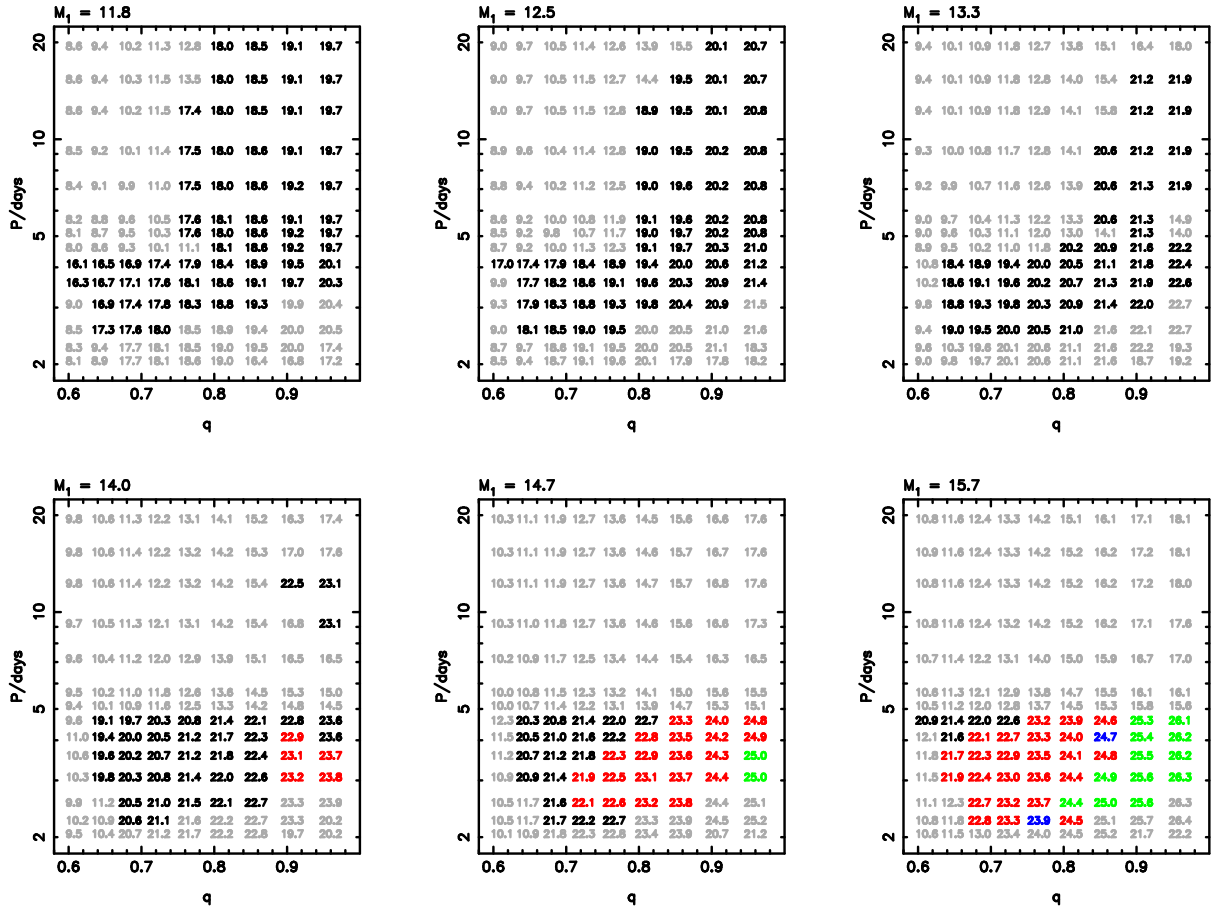,angle=270,width=160mm}
\caption{As Fig. 4 but for varying initial mass at solar metallicity. Although the number of stable case-B mass transfer systems decreases with increasing mass, until by $15 \msun$ none remain, the number of stable case-A systems increases.}
\vfil}
\label{fig5}
\end{figure*}

\subsection{Metallicity and enhanced-metallicity rejuvenation}
Metallicity affects the evolution of a binary in several ways. First, the radius of a lower-metallicity star is generally smaller 
at the equivalent evolutionary stage -- so a lower-metallicity binary avoids RLOF for longer. This means that, for example, the initial 
period boundary between case-A and case-B mass transfer is lower at lower metallicity. For the metallicity 
range ($Z = 0.03$ -- $0.0001$) shown in Fig. 4, this boundary decreases from over 5 days to around 2~days -- a 
significant drop, because at least at the higher metallicities over this mass range it is only case-A systems which 
avoid common envelope evolution. However, a second effect of the radius decrease at low metallicity is
that the secondary star also has a smaller radius, thus rendering some 
case-B systems stable against contact which at higher metallicities would not have been. In fact, the 
behaviour in terms of contact/non-contact systems (see Pols 1994; Wellstein et al. 2001; Dray \& Tout 2005)
with decreasing metallicity is similar to that with decreasing mass (Fig. 5) after the differences in the case-A/case-B
boundary are accounted for. The smaller radius of the secondary also affects short-period case-A systems. As described in 
Wellstein et al. (2001), there is a class of systems with short initial periods which evolve in a contact-free manner 
through an initial case-A mass-transfer phase but, because this happens very early on in the lifetime of the binary, the 
secondary's evolution then overtakes that of the primary and it attempts a later phase of reverse case-B mass transfer whilst 
the primary is still in its core hydrogen burning phase and this leads to contact. These systems lie under the main area of 
contact-free evolution in Figs. 4 and 5 and can be identified by their high secondary masses at the time of contact. Because the 
smaller radii at lower metallicities delay the onset of mass transfer, more systems at low metallicity avoid this fate and 
hence the area of contact-free evolution extends downward in period significantly. The combination of these effects leads to 
there being virtually no overlap at all between the parameter space of contact-free case-A systems at metallicity 0.03 and 
metallicity 0.0001. A further effect, mainly important for the higher-mass end of the parameter space we look at here, is that 
line-driven wind mass-loss rates are lower at lower metallicity. As less mass is lost in the wind of the primary, more mass 
is potentially available for transfer to the secondary, and the angular momentum lost this way also affects the period evolution. 

The effects of enhanced-metallicity rejuvenation are apparent in Figs. 4 and 5 from the final subtype distribution of the 
contact-free secondaries (indicated by colour\footnote{It should be noted that the criteria used to classify a star as WR or not include that its effective temperature obey $\log_{10}(T_{e}/K) > 4.0$. This, combined with the sometimes rather sensitive response of evolutionary tracks to 
small increases in mass loss, is what leads to the occasionally ragged boundary between areas of the parameter space in which 
contact-free secondary evolution results in the various WR subtypes.}). At solar metallicity, single star 
models with the same physical ingredients
as the binary models used here do not go through a WR phase unless they are initially over around $28\,\msun$. However, 
an initially $14 + 12.5\,\msun$, ${P = 3}$ days binary -- that is, one with a {\it combined} mass which is lower than 
the single star WR-forming limit -- produces a secondary which goes through a WN phase. Given that the primary, after its 
envelope is stripped by RLOF, can also appear as a WN or WN-like star (although this phase is unlikely to be concurrent with the 
WR phase of the secondary), this is very much a two for the price of none channel for WR production! However, the stripped primary 
is usually undermassive for a WR star and may be underluminous, so it is uncertain whether it would be observable as one. In contrast, 
accreting secondaries are overluminous and can go through both WN and WC phases. As is shown in Fig. 6, 
the effect of enhanced-metallicity rejuvenation is similar, although not identical, to a straightforward upwards shift in 
metallicity over the range for which we have detailed coverage of the transition region. In fact Fig. 6 is not quite 
comparing like with like. We are showing the {\it initial} masses for WR formation for single stars against the {\it maximum}
masses of secondaries which become WR stars. At the point of maximum mass, the secondaries are not ZAMS stars; the closest single 
star equivalent is one which is partway through the main sequence and has already (especially if it is of high metallicity) 
lost some mass. For example, the 
closest match to the evolution of the solar-metallicity secondary which forms the lower accretion star WN mass limit 
shown in Fig. 6 is a single star with initial mass closer to $23\,\msun$, offsetting the limit upwards by just over half 
a solar mass. In fact, there is not one single maximum-mass limit for a secondary to undergo WR evolution at a particular 
metallicity -- whether it does or not depends on the amount of helium accreted, the amount of helium synthesised in 
the secondary's core before the onset of mass transfer and the time of mass transfer (e.g. Braun \& Langer 1995)
both of which are determined by the initial binary parameters. Thus there is 
some variation in the threshold mass -- however, in practice, this variation is only around a solar mass for fully-conservative systems.

\subsection{Approximate formulae}

Whilst the formulae used for the partial-accretion model above work well in some cases, they do not work so well for all models.
In particular, models for which the metallicity and time offset behaviour are most successfully-matched are generally those with 
periods roughly in the middle of the case A range and are at the bottom or middle of the mass range. Differences are partly systematic, 
with the predicted time offset being too large for long periods by up to 30 \% and occasionally too small for short periods. Similarly, 
the metallicity of the best-fit model and the helium metallicity sometimes differ by a small amount, although 
this does not appear to be systematic. This is likely to be partially an 
effect of our fitting against a library of discrete models, particularly with regard to the metallicity (as the single star model grid is 
more closely-spaced in mass that in metallicity, for which we have only ten values available).
Of course, it should be noted once more 
that fitting to single stars can only ever be an approximation and that, at the higher end of the mass range the accretor may have 
synthesised significant amounts of helium itself before accretion occurs, so if it is thoroughly mixed a higher effective metallicity 
is expected. Also, as noted previously, there remains a dichotomy between fitting the HR diagram position with time well and fitting the mass 
evolution with time well -- both are possible, but frequently not with the same model (Fig. 3). With all this noted, the results of 
fitting using equations 1 and 2 with the metallicity-dependent main-sequence lifetime equation from Hurley et al. (2000) are still 
significantly better than assuming standard rejuvenation at the same metallicity.

\begin{figure}
\vbox to91mm{\vfil
\psfig{figure=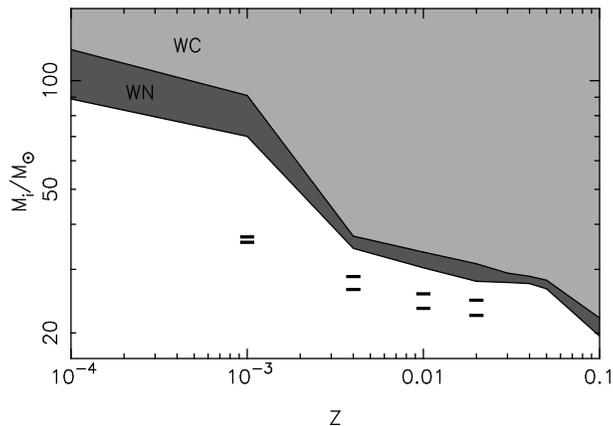,angle=270,width=80mm}
\caption{Mass limits for stars to end their lives in WN and WC sub-phases for single star models with the same 
physical ingredients as the binary models. Note that the criterion for WR evolution used here (surface hydrogen 
abundance below $0.4$ by mass and $\log_{10}(T_{\rm e}/K) > 4$) is less meaningful at the highest metallicity shown here, 
because the low initial hydrogen abundance makes the production of stars that fulfil it rather easy. Also shown 
are the lowest maximum-mass limits for stars to end their lives in the WC and WN phases (upper and lower thick black 
lines) for those metallicities at which we have performed a complete parameter study of the region in which the
transition occurs.}
\vfil}
\label{fig6}
\end{figure}

\section{Discussion}

The effects of conservative accretion which we have looked at in this paper can manifest observationally in a 
number of ways. For instance, if at least some mass transfer is helium-enriched, we would expect a corresponding 
increase in the type Ibc/type II supernova rate ratio, the (runaway) WR star population and the CNO production 
from massive stars. What is not clear is whether such effects can be distinguishable against the uncertainties
arising from the question of how much matter can be accreted in any given binary.

If one assumes that primaries of binaries have a Salpeter IMF, mass transfer is conservative, the initial mass ratio 
distribution of binary systems is flat\footnote{A mass ratio distribution weighted towards equal-mass or twin systems 
gives very similar results.} and the initial period distribution is flat in $\log(P)$, the inclusion of 
enhanced-metallicity rejuvenation in initially solar-metallicity accretion stars 
increases the population of secondaries which go through a WR phase by about 30 \%
(excluding those systems which survive common-envelope evolution). At the metallicity of the Small Magellanic Cloud the population is nearly doubled. 
This suggests that this is potentially an observable effect, if it can be distinguished from the other uncertainties which 
beset binary evolution. Since WR stars produce much larger amounts of carbon and other elements in their winds than stars which 
do not go through a WR phase, this population increase may have implications for the enrichment from WR stars (e.g. Dray \& Tout 2003), 
even though stars in binaries which avoid a contact phase and go on to become WR-like are a relatively small part of the 
total population. It is also worth considering what happens to binaries which interact and do not manage to avoid contact. If the 
outcome is a period of common-envelope evolution during which the secondary accretes no further matter, followed by further evolution
as a detached but close binary, then we would expect the evolution of the secondary to continue as discussed in section 3 and 
enhanced-metallicity evolution would have a negligible effect in most cases. If instead the components of the binary merge 
as a consequence of this process -- a fate which may happen to a majority of interacting binaries -- then it is likely that 
the structure of the resulting star will have been quite thoroughly mixed and it would behave similarly to the conservatively-accreting 
secondaries previously discussed. If this happens, the population of binary-formed WR stars may be much larger than estimated above. 
It is also interesting to note that a large proportion of these binary-formed WR stars would be single during their WR phase.

However, whether or not it can fit the whole-population properties, a successful model of massive-binary evolution must also 
be able to explain individual systems. In particular, the properties of individual systems may be used to constrain 
the model set used because, with the large number of free or poorly-constrained parameters available to vary in population synthesis, 
it is not hard to make most of the possible populations fit in some way or other. 
The rejuvenatory effects discussed in this paper only come into effect if there is either
conservative of nearly-conservative mass transfer or if there is a late mass transfer event (not necessarily wholly conservative) in which 
matter from a star with an already strongly helium-enriched surface composition is transferred across. As discussed above, if only the first 
ten percent or so of matter in a mass transfer event can be accreted (e.g. Packet 1981; Dewi 2006; but see also Petrovic et al. 2005, in which 
mass transfer is linked to rotation in such a way that an initial mass transfer event can be  non-conservative but a later event more 
nearly conservative) then the potential metallicity 
gain-like effects of accretion would be basically negligible. However, if some proportion of systems do accrete a large amount of 
helium-enriched matter, then this could produce potentially observable effects, at least in individual systems. Thus observation of these
effects could potentially be used to provide some constraint on the amount of matter accreted in a binary.

However, as noted previously, it is probable that the amount of matter which can be accreted is variable and affected by a wide 
range of other parameters, not least rotation (Langer et al. 2003). 
In our previous work modelling binary populations (with slightly less detailed models; 
see e.g. Dray \& Tout 2006), based on the 20 known WR binaries with measured masses in the catalogue of van der Hucht (2001)
we found that no particular accretion scenario is favoured for all stars -- most systems which are easy to fit if we assume
conservative evolution are also easy to fit using non-conservative evolution but different starting parameters. Some 
systems require nearly-conservative evolution (e.g. Dray 2006) and some require non-conservative evolution.
There is a further group of binaries whose period is 
too small to fit well (and which are therefore presumably post-common envelope systems, whose evolution we have not followed
in detail due to the large uncertainty surrounding common envelope evolution) and a couple of unusual systems 
whose origin may be dynamical. Some dynamically-formed systems are likely because some stars have exchanged companions in dense environments in 
the past (Vanbeveren 2005). This may also be the case in those systems which contain two vary massive stars, such 
as WR20a and WR47, although these are expected if the initial binary distribution is skewed towards twin systems 
in which both stars start with similar masses (Pinsonneault \& Stanek 2006), perhaps as a natural result of the mode of 
massive star formation (Bonnell \& Bate 2006).
Investigations of whole-population properties with these models (Dray et al. 2005; Dray 2006) again suggested a picture in which 
some mass transfer is conservative and some not. In this rather complex scenario, the uncertainty associated with mass transfer 
is so much greater than 
that associated with the precise effects of rejuvenation that not only is it difficult to say anything concrete about the latter from comparison 
of whole-population properties with models, but constraints on the mass accreted in an individual binary, if not coupled to 
a detailed understanding of that binary's previous evolution, may say little about the mass transfer situation in general other 
that providing a rather weak limit. 

It may still be possible to observe the effects of enhanced-metallicity rejuvenation in individual stars and groups of stars. In 
particular, we expect from our models that these effects would be most visible in the time period after the SN explosion of the primary
in a system which has been undergoing mass transfer (if the SN order is not reversed -- see e.g. Pols 1994). If the SN does not unbind 
the system, then the result would be a massive star -- compact object binary. Cyg X--3 is an example of such a system in which there 
is a WR star; however, it is very unusual (Lommen et al. 2005). The most common fate, as discussed in Dray et al. (2005) is that the 
system is split by the SN kick, and the potentially-rejuvenated component becomes a high-velocity runaway star. Many runaway O stars have 
been observed to have enhanced surface helium abundances (Blaauw 1993), as found in our models. A notable subgroup of 
the WR population is the WN8 stars, which differ significantly from normal WR stars (Marchenko et al. 1998). Many of these stars are 
measurable runaways or appear to have moved a significant distance from their place of birth and very few are in binaries. Whilst 
it has been suggested that they could be the result of massive star--compact object mergers (i.e. a type of Thorne-{\.Z}ytkow object, 
Vanbeveren et al. 1998; Cherepashchuk \& Moffat 1994), it is quite possible that they are just the unusual WR stars which result from 
secondary evolution in an accreting binary. Tracking of these stars and of runaway O stars back to their places of origin and a 
comparison of the metallicity of that area with their apparent metallicity from massive star evolutionary tracks and surface compositions could 
potentially give a limit on how important enhanced-metallicity rejuvenation is, and whether it is an effect that needs to be included 
in population synthesis models or not.

\section*{Acknowledgements}

LMD is grateful for support by the Leicester PPARC rolling grant for
theoretical astrophysics, and to Peter Eggleton for providing and 
explaining his code. CAT thanks Churchill College for his Fellowship.

\end{document}